\magnification=\magstep1
\advance\mathsurround2pt
\font\ninerm=cmr9

\headline={\it J. J. Lodder, Preprint. \hfill
hep-ph/9606306
\hfill
Submitted to {\rm Nucl. Phys. B}}

\centerline{\bf\uppercase{On Electro-weak mixing }}
\medskip
\centerline{\bf\uppercase{derived from radiative corrections}}
\medskip
\centerline{\bf\uppercase{and the necessity of the Higgs mechanism}}
\medskip
\bigskip
\centerline{ J. J. LODDER}
\smallskip
\centerline{\it Oudegracht 331\thinspace b,
3511\thinspace\thinspace PC Utrecht,
The Netherlands.\rm \footnote*{E-mail: jjl@knoware.nl}}
\bigskip\bigskip\bigskip

{\ninerm\noindent
{\bf Abstract} Starting from the unmixed
Lagrangian,
the electro-weak radiative corrections are recomputed using
symmetrical generalised functions.
All gauge bosons acquire an indeterminate mass.
Electro-weak mixing is obtained by
diagonalization of the mass matrix.
The W/Z mass ratio follows automatically.
The Higgs mechanism is not needed to generate the vector boson masses.
A mass sum rule
is postulated to obtain a zero photon mass.
The radiative corrections are different,
and (provisionally) in agreement with experiment,
when the no longer needed Higgs terms are omitted.}
%
%
\begingroup
\def\sss{\scriptscriptstyle}
\newskip\diaskip \diaskip=9mm
\newfam\bfsyfam

\font\tenbfsy =cmbsy10
\font\sevenbfsy=cmsy7
\font\fivebfsy =cmsy5
\textfont\bfsyfam =\tenbfsy
\scriptfont\bfsyfam =\sevenbfsy
\scriptscriptfont\bfsyfam=\fivebfsy
\skewchar\tenbfsy='60 \skewchar\sevenbfsy='60 \skewchar\fivebfsy='60
\newfam\mitbfam
\def\mitb{\fam\mitbfam}
\font\tenmitb =cmmib10
\font\sevenmitb =cmmi7
\font\fivemitb =cmmi5
\textfont\mitbfam  =\tenmitb
\scriptfont\mitbfam =\sevenmitb
\scriptscriptfont\mitbfam =\fivemitb
\skewchar\tenmitb='177\skewchar\sevenmitb='177\skewchar\fivemitb='177
\catcode`@=11
\def\hexnumber@#1{\ifnum#1<10 \number#1\else
 \ifnum#1=10 A\else\ifnum#1=11 B\else\ifnum#1=12 C\else
 \ifnum#1=13 D\else\ifnum#1=14 E\else\ifnum#1=15 F\fi\fi\fi\fi\fi\fi\fi}
\def\bffam@ {\hexnumber@ \bffam}
\def\mitbfam@ {\hexnumber@ \mitbfam}
\def\bfsyfam@ {\hexnumber@ \bfsyfam}
\def\msymfam@ {\hexnumber@ \msymfam}
\mathchardef\taubf = "0\mitbfam@1C
\mathchardef\cdotbf = "2\bfsyfam@01
\catcode`@=12
\def\lover#1/#2 {{\textstyle{#1\over#2}}}
\def\eq(#1){(#1)}
\def\ref#1{ [#1]}
\def\Tr{\mathop{\hbox{\rm Tr}}}
\def\intdppi{\int\!\!{d^{\,4}p\over(2\pi)^4}\,\,\,}
\def\intdp{\int\!\!d^{\,4}p\,\,\,}
\def\msq{m^2}\def\mt{m_{t}^{\vphantom2}}
\def\mtsq{m_{t}^{\,2}}

\def\mwsq{m_{\rm\sss W}^{\,\,2}}
\def\mz{m_{\sss \!Z}^{\vphantom2}}
\def\mzsq{m_{\sss \!Z}^{\,\,2}}
\def\mthreesq{m_{\sss W^3}^{\,2}}
\def\lmt{\log(\mtsq)}
\def\mm{{\mitb M}}
\def\psq{p^2}
\def\phat{\hbox{$\not\mkern-3.3mu p\mkern3.3mu$}}
\def\cv{c_{\sss V}}
\def\ca{c_{\!\sss A}}
\def\cvsq{\cv^{\,2}}
\def\casq{\ca^{\,2}}
\def\gw{g_w}
\def\gp{{g^{\sss\mkern 1 mu \prime}}}
\def\ge{g_e}
\def\gz{g_{\sss Z}^{\hphantom2}}
\def\gwsq{\gw{}^{\!\!2}}
\def\gpsq{\gp^{2}}
\def\gesq{\ge{}^{\!\!2}}
\def\gzsq{\gz{}^{\!\!2}}

\def\C{{\it C}}
\def\thw{\theta_w^{\phantom2}}
\def\cwsq{\cos^{2}\!\thw}
\def\cwqu{\cos^{4}\!\thw}
\def\swsq{\sin^{2}\!\thw}
\def\swqu{\sin^{4}\!\thw}
\def\gmnu{g^{\mu\nu}}
\def\gmnd{g_{\mu\nu}}
\def\gfiveu{{\gamma^5_{\vphantom2}}}
\def\gfived{{\gamma_{5}^{\vphantom2}}}
\def\drho{\Delta\rho}
\def\pimm{{\Pi}_\mu^{\,\mu}(0)}
\def\gev{\hbox{\rm GeV}}
\def\wpm{W^\pm}
\bigskip
\noindent
{\bf 1. Introduction}
\bigskip
\noindent
A new theory of generalised functions has been constructed\ref{1,2}.
The available simple model allows the multiplication
of all generalised functions needed
in quantum field theory.
Standard concepts of analysis,
such as limit,
derivative,
and integral,
have to be extended to make multiplication of generalised functions possible.
Integration between arbitrary limits is always possible and yields a
well-defined finite result.
Infinity of integrals is replaced by the less restricted concept
of determinacy,
which is related to the scale transformation properties of the integrand.
In contrast to all regularization schemes the results
{\it are not arbitrary
by finite renormalizations}\/\ref3.

Conversely,
all results in quantum field theory that depend on the use
of {\it any particular method of regularization}\/ are invalid
by the standard of the symmetrical theory of generalised functions.
In particular,
every result that is dependent on the use of dimensional regularization
disagrees with the corresponding generalised function result\ref3.
This is not merely a mathematical nicety,
it has physical consequences for observable quantities.
The usual computations of the radiative corrections in the standard model
found in the literature\ref4 are an example\ref5.

One should define one's mathematical tools before one starts calculating,
instead of adjusting the definitions in order to obtain results
supposed to be desirable.
The added strength of the generalised function method comes from its
requirement that all of mathematical analysis should be constructed in
such a way that multiplication of generalised functions is possible,
instead of fixing things up afterwards when trouble occurs.

In this paper we calculate the second order vacuum polarization
in the unmixed standard model with the Higgs mechanism omitted.
The gauge symmetry breaks automatically,
and electro-weak mixing occurs inevitably,
with the correct results.
The idea of dynamical symmetry breaking is not new.
In fact it was considered already\ref{6,7}
before the standard model became well-established.
It did not meet with success however.
This is understandable,
using renormalization methods the necessary mathematical tools are lacking.
The absence of arbitrary finite renormalizations
and the irrelevance of renormalizability are essential for this purpose.

\bigskip\goodbreak
\noindent
{\bf 2. The Lagrangian}
\bigskip
\noindent
The starting point is the usual~$SU(2)\otimes U(1)$ Lagrangian,
omitting the Higgs terms
$$\eqalign{
{\L}&=-\lover1/4 {W}^{\mu\nu}\!\cdotbf{W}_{\!\mu\nu}-
\lover1/4 {B}^{\mu\nu}{B}_{\mu\nu}+\cr
&\quad + i\bar \psi_L\gamma^\mu(\partial_\mu +
i\gw{\mitb W}_{\!\mu} \cdotbf{\taubf\over2})\psi_L+
i\bar\psi\gamma^\mu(\partial_\mu+i{\gp\over2}Y^{\sss W} B_\mu)\psi +
i\mt\bar tt\,,\cr}\eqno{(1)}$$
with a vector triplet coupled with strength~$\gw$ and an isosinglet~$B$
coupled to the weak hypercharge current~$Y^{\sss W} =2Q-2T^3$ with
strength~$\gp/2$,
and with the field strengths given by
$$\eqalign{
{\mitb W}_{\!\mu\nu}& =
\partial_\mu{\mitb W}_{\!\nu} - \partial_\nu {\mitb W}_{\!\mu} -
\gw {\mitb W}_{\!\mu}\times {\mitb W}_{\!\nu}\,,\cr
B_{\mu\nu}& =
\partial_\mu B_\nu - \partial_\nu B_\mu.\,\cr}\eqno{(2)}$$
The~$SU(2)$ symmetry is broken by hand.
The top quark is given a non-zero mass,
leaving the bottom quark (effectively) massless.
The leptons and other generations do not contribute,

\noindent
Note:
It is inevitable that one fermion mass is introduced to set the mass scale.
Gauge boson masses can result from radiative corrections,
conversely fermion masses cannot be generated
by interaction with massive gauge bosons.

We have to compute the charged~$W^{1,2}$ or~$\wpm$ radiative corrections
caused by the~$t\bar b$-quark loop,
and the neutral~$W^3$ and~$B$ radiative corrections
from virtual~$t\bar t$ pairs.
In addition the ~$W^3$ and the~$B$ mix,
since a~$W^3$ may create a~$ t\bar t$ pair,
which successively annihilates to a~$B$,
and visa versa,
giving rise to three related diagrams differing only by vertex factors.
It is convenient to calculate the loops
with vertex factor~$\cv\gamma^\mu-\ca\gamma^\mu\gfiveu$
and specialize afterwards.

The cubic
and quartic terms in~$W_\mu$,
from~${W}^{\mu\nu}\!\cdotbf{W}_{\!\mu\nu}$,
give rise to a W-boson self-interaction.
This is also needed to obtain physically correct radiative corrections.

Since only the mass terms have to be computed
it is not necessary to consider ghosts.
We can compute conveniently in the unitary gauge,
which is free of unphysical fields.

\bigskip\goodbreak
\noindent
{\bf 3. Charged vector bosons}
\bigskip
\noindent
The fermionic contribution to the~$\wpm$ self-energy is
found from the boson-fermion vertex and the
corresponding fermion loop diagram,
\vskip\diaskip
\vbox to 0pt{\vss\line{\hfill {Feynman diagram to be added}\quad \hfill}}
\vskip\diaskip
\noindent
which is needed for the present purpose only at boson
momentum~$k=0$.
Substitution of the
Feynman rules with the general vertex factor~\eq(A3) gives
$$\Pi_{\mu\nu}(0) =
-{3g^2\over4 }\Tr\!\intdppi\,\gamma_\mu(\cv-\ca\gfived)
{\phat+m_1\over p^2-m_1^{\,2}}\gamma_\nu(\cv-\ca\gfived)
{\phat+m_2\over p^2-m_2^{\,2}},
\eqno{(3)}$$
multiplied by an additional factor three for summing over the quark colours.
After evaluating the trace,
contracting with~$g^{\mu\nu}\!\!$,
and combining the denominators
with the usual Feynman trick,
one obtains
$$\pimm = {3g^2\over32\pi^4}\int^1_0\!\!\!dx\!
\intdp{(\casq+\cvsq)\psq+2(\casq-\cvsq)m_1m_2\over
(\psq-xm_1^{\,2}-(1-x)m_2^{\,2})^2}.
\eqno{(4)}$$
For~$\wpm$-bosons the dominant contribution is obtained by
taking the fermions to be a top and a bottom quark.
To sufficient accuracy the bottom quark and all leptons
are massless for the present purpose.

Substitution of~$g=\gw/\sqrt2$,
$\cv=\ca=1$,
$m_1=\mt$,~$m_2=m_b=0$,
and the value of the integrals~\eq(A1,2) from the appendix,
yields
$$\mwsq = \lover 1/4i \pimm_{\sss \wpm}^{\vphantom2} =
{}-{3\gwsq\over32\pi^2}\mtsq(\log\mtsq+\C-1).
\eqno{(5)}$$
It is seen from the appearance of an indeterminate constant~$\C$
that the result is indeterminate.
As usual, this means that the mass has to be supplied by
experiment.
 Fermion loops always give rise to indeterminacy,
except for the special case of vector coupling with~$m_1=m_2$.

\noindent
Note: It is not necessary to write~$\Delta\mwsq$ here,
since the W-bosons are massless
to begin with by gauge invariance.
The physical~$\wpm$ mass comes entirely
from the interaction with the fermion loop.
This means we have skipped an order:
The zero$^{\rm \!th\!}$ order mass has been generated
from second order perturbation
theory.

\bigskip\goodbreak
\noindent
{\bf 4. Neutral gauge bosons}
\bigskip
\noindent

The top/antitop vertex can couple left and right to a~$W^3$ or a~$B$,
giving rise to three different diagrams.
(The two mixed off-diagonal diagrams are equal.)
One calculation gives all three by specializing the coupling constants
and~$\ca$ and~$\cv$ appropriately.
The calculation proceeds as above,
with~$m_1=m_2=\mt$ equal.
The Feynman trick is not needed and the integrals evaluate directly to
$$\vcenter{\tabskip=0pt plus1fill \openup 4 \jot
\halign{$\displaystyle#$&$\displaystyle#$&$\displaystyle#$\cr
B\cdotbf{\bar t\atop t}&{}\cdotbf B\to\qquad\quad
m_{\sss B}^{\,2}=\lover1/4i \pimm_{\sss
BB}^{\vphantom2} & =
{}-{3\gpsq\over32\pi^2}\mtsq
(\lmt+\C-\lover17/18 ),\cr
B\cdotbf{\bar t\atop t}&{}\cdotbf W^3\to\quad
m_{\sss BW^3}^{2} =
\lover1/4i \pimm_{\sss BW^3}^{\vphantom2} & =
{}+{3\gp\!\gw\over32\pi^2}
\mtsq(\lmt+\C+ \lover1/6 ),\cr
W^3\cdotbf{\bar t\atop t}&{}\cdotbf W^3\to\quad
\mthreesq=\lover1/4i \pimm_{\sss W^3W^3}^{\vphantom2}& =
{}-{3\gwsq\over32\pi^2}
\mtsq(\lmt + \C - \lover1/2 ).\cr }}\eqno{(6)}$$

It should be noted that the numbers appearing
with the indeterminate constant~$\C$
are by themselves meaningless,
since any finite number can be added at will.
Nevertheless,
by keeping them consistently,
one can compute determinate differences
of indeterminate quantities\ref{1-3}.

\bigskip\goodbreak
\noindent
{\bf 5. Electro-weak mixing}
\bigskip
\noindent
The vacuum polarization terms contribute a quadratic form
in~$(B,W^3)$ to the effective second order Lagrangian,
characterized by a (squared) mass
matrix~$\mm$.
Collecting all results gives the second
order mass matrix for the neutral vector bosons
$$\mm=
{}-{3\mtsq\over32\pi^2}\pmatrix{
\hphantom{-}\gpsq(\C-\lover17/18 ) &
-\gp\gw(\C+\lover1/6 \vphantom{\bigg(})\cr
-\gp\gw(\C+\lover1/6 ) &
 \vphantom{\bigg(}\hphantom{-}\gwsq(\C - \lover1/2
)\cr}\eqno{(7)}$$
(Temporarily omitting the $\lmt$ for clarity.)
The~$\C^2$ terms cancel in the determinant of~$\mm$,
so only one of the~$\C$-eigenvalues can be indeterminate.
The indeterminate part of~$\mm$ is diagonalized by the transformation
$$\eqalign{
A_\mu & = +B_\mu\cos\thw + W_\mu^3\sin\thw\,,\cr
Z_\mu & = -B_\mu\sin\thw + W_\mu^3\cos\thw\,,\cr}
\eqno{(8)}$$
with the conventional definitions
$$\tan\thw={\gp\over \gw}\qquad \hbox{\rm so} \qquad\sin\thw={\gp\over
\gz}\qquad\hbox{\rm with}\qquad\gzsq=\gwsq+\gpsq.\eqno{(9)}$$
The neutral mass matrix is transformed by the electro-weak rotation into
$$\mm = {3\mtsq\over32\pi^2}\pmatrix{
\gesq\lover16/9 \vphantom{\bigg(} &
\ge\gz(\lover2/3 -\lover 16/9 \swsq) \cr
\ge\gz(\lover2/3 -\lover 16/9 \swsq ) &
{}-\gzsq(\C+\lover1/2 -\lover4/3 \swsq +\lover16/9
\swqu)\vphantom{\bigg(}\cr}\eqno{(10)}$$
By definition the indeterminate eigenvalue is assigned to the~$Z_\mu$.
It can be put equal to the experimental mass,
as usual in quantum field theory.

Comparing the~$Z$-mass and
the previously found~$W$-mass~\eq(5),
one sees that both are indeterminate,
but ignoring finite terms for the time being,
their ratio is determinate
$${\mwsq \over \mzsq} = {\gwsq\over \gzsq} = \cwsq +
{\cal O}\left(\gwsq{\mtsq\over\mwsq}\right).
\eqno{(11)}$$
The radiative corrections to this equation,
which have been computed already in~\eq(5) and~\eq(6)
will be compared in section~8 with the standard model results.

The second order effective Lagrangian almost agrees with the standard model
Lagrangian with mass terms after the Higgs rotation.
It has the correct couplings and masses.

The~$A^{\mu\!}A_\mu$ element of the mass matrix
must be interpreted as a finite,
non-zero photon mass,
in apparent disagreement with
electromagnetic gauge invariance.
The result agrees with a previous calculation\ref8 based on QED alone.
It is determinate,
so it must have a physical meaning.

However,
the initially massless~$W$ and~$Z$ vector bosons now have acquired a
mass,
so it is necessary to consider also the boson self-interactions introduced by
the non-Abelian group.
These will contribute self-mass terms,
which must
be taken into account in order to be consistent.
This peculiar situation is a consequence of the initial masslessness
of the gauge bosons.

\bigskip\goodbreak
\noindent
{\bf 6. Vector boson self-interaction}
\bigskip
\noindent
The cubic and quartic terms in the free field part of the Lagrangian
give rise to vector boson self-interaction.
Massless vector bosons do not have a mass term in the self-interaction.
When the vector bosons have acquired a mass
(by any mechanism) there will be
self-mass corrections.
All self-interactions have the same structure,
so we compute only need to compute the generic case,
for example the ~$W^3$ self-mass
The others are equal to it by cyclic permutation.

There are two Feynman diagrams contributing to the self-energy
\vskip\diaskip
\vbox to 0pt{\vss\line{\hfill {Feynman diagram to be added}\quad\hfill}}
\vskip\diaskip
\noindent
obtained from the vertices~\eq({\bf A}8,9).
The 3-boson vertices give a loop diagram.
One finds that the terms involving~$p^4p_\mu p_\nu $ cancel,
leaving upon contraction with~$\gmnu$ the quartic and quadratic terms
$${\pimm}_{\!\rm\sss loop}^{\phantom2}=
{}-6{g^2\over\msq}
\intdppi{p^4-3\psq\msq\over(\psq-\msq)^2},\eqno{(12)}$$
The 4-boson vertex~\eq({\bf A}9) gives the bubble diagram,
which yields upon contraction with a propagator
$${\pimm}_{\!\sss\rm bubble}^{\phantom2}=
{}+6{g^2\over\msq}
\intdppi{\psq-4\msq\over\psq-\msq}.\eqno{(13)}$$
The quartic terms cancel in the sum of the two diagrams,
so the total vector boson self-mass is
$${\Delta\msq}_{\!\!\! \sss\rm self}^{\vphantom2}=
\lover 1/4i {\pimm}_{\!\rm\sss total}^{\phantom2}=
{}-{3g^2\over 16\pi^2}
\intdp{\psq-2\msq\over(\psq-\msq)^2}
={3g^2\msq\over16\pi^2},
\eqno{(14)}$$
which is the same determinate integral found previously~\ref3
for the photon mass.
One sees that determinacy and degree of divergence are not related.
Substituting the value of the integral,
the final result is a determinate self-mass correction
$$
\Delta\msq{}_{\!\!\!\sss\rm self}^{\vphantom2}=
{}-{3g^2\over16\pi^2}\msq.\eqno{(15)}$$
This term is absent when dimensional regularization is used to evaluate the
integral.

The additional~$W^3W^3$ element of
the neutral mass matrix transforms under the electro-weak rotation into
$$\Delta\mm_{\rm self} =
{}-{3\mwsq\over16\pi^2}
\pmatrix{\gesq & \ge\gz\cwsq \cr
\ge\gz\cwsq & \gzsq\cwqu\cr}\eqno{(16)}$$
The self-interaction gives determinate contributions
to all mass matrix elements found above.

There is also a self-correction to the~$\wpm$ mass.
It can be calculated in the same way,
from the same Feynman diagrams
with the result
$${\Delta\mwsq}_{
\sss\rm self}^{\vphantom2}={}-{3g^2\mwsq\over16\pi^2}.
\eqno{(17)}$$
The~$ZZ$ and~$\wpm \wpm$
self-terms contribute to the radiative corrections of
the~$W/Z$ mass ratio.
One can also find the same results by computing the vacuum polarization
with the standard model Feynman rules.

The self-interaction gives a determinate extra photon mass term,
and one sees that the fermionic and bosonic contributions
to the photon mass have opposite sign,
so they may cancel.

\bigskip\goodbreak
\noindent
{\bf 7. The mass sum rule}
\bigskip
\noindent
The determinate fermion and boson contributions to the photon mass have
opposite sign.
This may be seen as a consequence of the additional minus sign
associated with a fermion loop.
The photon mass will be zero when
$$\mtsq=\lover9/8 \mwsq+
\hbox{other boson and fermion contributions},\eqno{(18)}$$
as found before\ref8 on basis of QED alone.
The physical implications of this equation,
when it is combined with the assumption
of completeness of the standard model,
will be discussed elsewhere\ref9.

The off-diagonal elements of the mass matrix
do not cancel with the photon mass,
so we are left with a
two-particle~$A_\mu\leftrightarrow Z_\mu$ counterterm.
This does not cause any problems since it is determinate.
It cannot be helped while we do not possess a unified account
of the electro-weak interaction.

Equation~\eq(18) is a special case of the more general
boson/fermion mass sum rule
derived previously\ref8 from QED alone.
$$\sum_{\hbox to 0pt{\sevenrm\hss fermions\hss}}
\,g_f^{\phantom{2}}q_{\!f}^{\,2}c_jm_{\!f}^{\,2}=
\sum_{\hbox to 0pt{\sevenrm \hss bosons\hss}}
\,g_{b}^{\phantom{2}}q_b^{\,2}c_jm_b^{\,2},
\eqno{(19)}$$
The sums are to be taken over all fundamental fermions and bosons.
The factors~$g_f$ and~$g_b$ are the multiplicity of the
particles,
and the~$q^2$'s are the charges measured in units of
the squared electron charge.
The factors~$c_j$ are the relative coefficients
of the integrals obtained from the vacuum polarization diagrams,
to be calculated for each spin value.
One finds~$c_j=2J+1$,
for~$J\leq1$, \ref8.

It is interesting to see that meaningful cancellation of
bosonic and fermionic contributions to undesirable results
may occur without invoking supersymmetry.
There is a difference;
with supersymmetry one hopes for cancellation of undesirable infinities,
in the symmetrical theory of generalised functions
applied to quantum field theory there are no infinities.
Instead one has cancellation
of physical predictions in disagreement with experiment.

It remains to be seen if the cancellation can be proved to all orders.
There is good reason to believe that this is the case,
since the results of generalised function calculations differ
only by finite renormalizations from the results
obtained by any standard regularization method.
Any severe problems with the generalised function
approach should be present in the standard account as well.

A mass sum rule such as~\eq(19) is as yet no constraint on the theory,
as it can always be assumed to be satisfied.
It is dominated by the heaviest particles,
and one can always postulate as yet undiscovered,
very heavy,
new fundamental particles to close the sum rule.

Conversely,
when the mass sum rule is known to hold,
completeness of the physical theory may be surmised.
This may constrain as yet unavailable theories of everything in the future.
This point will be taken up elsewhere\ref9.

\bigskip\goodbreak
\noindent
{\bf 8. Radiative corrections and the existence of the Higgs boson}
\bigskip
\noindent
The radiative corrections computed by means of generalised functions
differ from the same results calculated in the standard manner.

The relation between the vector boson
masses and the weak mixing angle is not precisely given by~\eq(11).
It is modified by higher order radiative corrections.
It is customary to describe these
by means of the~$\rho$-parameter defined by
$$\rho\mathrel{:=}{\mwsq\over\mzsq\cwsq}.\eqno{(20)}$$
In the standard treatment the second order correction
to~$\rho$ is found to be \ref4
$$\drho_{\rm\sss standard}^{\phantom2}=
{1\over4\mzsq}\pimm_{ZZ}-{1\over4\mwsq}\pimm_{WW}=
{3\gwsq\over32\pi^2}{\mtsq\over\mwsq}.\eqno{(21)}$$
This expression has been evaluated using the standard Feynman rules,
and dimensional regularization has been used to evaluate the integrals.
Consequently only the axial part of the coupling contributes.
The vector part is proportional to the usual photon mass integral,
which is forced to be zero by the use of dimensional regularization.

In section~\eq(5) the finite terms were ignored.
Actually this is not the correct way to handle the indeterminacy.
Instead one has from~\eq(5) and~\eq(10)
$$\mzsq\!\cwsq -
\mwsq=
{3\gwsq\mtsq\over32\pi^2}
(1+{8\over3}\sin^2\thw-{32\over9}\sin^4\thw),\eqno{(22)}$$
which is determinate and in agreement
with the radiative correction\ref5 calculated before.
In terms of the~$\rho$-parameter\ref4 this translates to
$$\drho=\rho-1=
{\mwsq-\mzsq\cwsq\over\mzsq\cwsq}=
{3\gzsq\over32\pi^2}{\mtsq\over\mzsq}
(1+{8\over3}\sin^2\thw-{32\over9}\sin^4\thw).\eqno{(23)}$$
The result of the generalised function calculation differs
by a finite renormalization
from the usual result\ref4 obtained by dimensional regularization.
In the standard treatment one finds only the first term corresponding
to the axial part of the generalised function result.
Also the pre-factor contains~$\gzsq/\mzsq$ instead
of~$\gwsq/\mwsq$,
but that is equivalent to this order.

The self-interaction of the weak vector bosons also contributes to~$\drho$,
since it contributes finite terms to both the~$W$ and~$Z$ mass.
Substituting~\eq(16) and~\eq(17) one finds
$$\drho_{\rm self}^{\vphantom2}=
-{3\gwsq\over16\pi^2}(1-\cwqu).\eqno{(24)}$$
Combining both contributions one finds to second order
$$\drho={3\gwsq\over64\pi^2}\left({\mtsq\over\mwsq}+
\left({8\mtsq\over3\mwsq}-8\right)\swsq-
\left({32\mtsq\over9\mwsq}-4\right)\swqu\right).\eqno{(25)}$$
Finally,
the standard model has a Higgs contribution to~$\drho$.
It is needed to obtain agreement with experiment,
since it is the only negative contribution to~$\drho$ in the standard model.
It has been evaluated to\ref4
$$\drho_{\sss\rm H}^{\vphantom2}
=-{3\gwsq\swsq\over32\pi^2\cwsq}
\ln\left(m_{\sss H}\over\mz\right).\eqno{(26)}$$
In the generalised function treatment the self-interaction
of the~$W$- and~$Z$-bosons also gives a negative~$\drho$.
Comparing the extra terms with the Higgs correction one finds the
equivalent, to be added, Higgs mass
$$\Delta m_{\sss H}=-\mz e^{1.9}\approx -600\,\gev,\eqno{(27)}$$
which is no longer needed or allowed.
The Higgs boson may be still have a mass close to~$\mz$,
but this seems to be excluded on other grounds.

The precission measurements of the radiative corrections,
which are interpreted as evidence for the existence of the Higgs boson,
may be reinterpreted as evidence for its non-existence
in the generalised function approach.
These limits have large experimental errors at present.
They will be strengthened considerably in
the near future when accurate measurements
of the~$W$ mass become available at LEP.

\noindent Note: These remarks are preliminary as yet.
The LEP experiments have been
interpreted using formula dependent on dimensional regularization.
The interpretation of
the experimental results may therefore contain second order errors.
It will be necessary to redo the analysis of the data,
using the generalised function results,
before firm conclusions can be drawn.
It is a challenge to see whether the LEP precission data
can be fitted to the theory without the Higgs mass,
so effectively with one free parameter less.

\bigskip\goodbreak
\noindent
{\bf 9. Conclusions}
\bigskip
\noindent
To lowest order the effective Lagrangian for the electro-weak sector
of the standard
model has been derived from the computation of radiative corrections.
The gauge boson masses appear naturally and inevitably,
with the correct mass ratio.
There is no need to postulate a Higgs mechanism.
The electro-weak Lagrangian appears with the standard model couplings.
The fermion masses remain arbitrary for the time being.

The recomputed radiative corrections probably leave no room
for a Higgs contribution,
but this will not be definite until the~$\wpm$ mass has been measured
to high precission.
The data analysis will have to be corrected for use of formulae
dependent on the use of dimensional regularization.

\bigskip\goodbreak
\noindent
{\bf Acknowledgement}
\medskip
\noindent
Dr.~H.~J.~de~Blank is thanked for critical discussions
and help with the computation.

\bigskip
\noindent
{\bf Appendix }
\bigskip
\noindent
Note: The integrals were computed in \ref3
with the convention~$\gmnd = (-,+,+,+)$.
The sign changes needed
for the modern convention~\hbox{$\gmnd = (+,-,-,-)$}
are easily added.

\noindent
The integrals we need are
$$\eqalignno{\intdp{1\over(\psq-a^2)^2}&=
-i\pi^2(\log a^2+\C),&{(\bf A} 1)\cr
\intdp{\psq\over(\psq-a^2)^2}&=
-i\pi^2a^2(2\log a^2+2\C-1),&{(\bf A} 2)\cr}$$
with~$\C$ the indeterminate constant,
and the~$i\epsilon$ in the denominator
understood.
Purists may read~$\log a^2$ as~$\log a^2/M^2$,
with~$M$ an arbitrary unit of mass,
but this does not influence any physical result.
The~$\C$ convention has the mnemonic advantage that the rules for
handling the~$\C$ are the same as those
of the~$\C$ in the indefinite integral.

\def\strut{\hbox to 0pt{\hss$\vphantom{\bigg(}$\hss}}
\noindent
The general vertex factor is used in the form
$$\hbox{\rm vertex} =
-i{g\over2}\gamma^\mu(\cv-\ca\gfiveu).
\eqno{({\bf A} 3)}$$
The unmixed Lagrangian vertex factors are
$$\def\tablerule{\noalign{\hrule}}
\def\-{\hphantom{-}}
\vcenter{\halign{
\vrule#&\strut\quad\hfill$#$\quad\hfill&\vrule#&
\quad\hfill\strut$#$\quad\hfill&
\vrule#&\strut\quad\hfill$#$\quad\hfill&\vrule#&
\strut\quad\hfill$#$\quad\hfill&\vrule#\cr
\tablerule
&\rm Particles && g && \cv &&\-\ca&\cr
\tablerule
&t\bar t B_\mu &&\gp&&\-\lover5/6 &&-\lover1/2 &\cr
\tablerule
&b\bar bB_\mu && \gp && -\lover1/6 &&\-\lover1/2 &\cr
\tablerule
&t\bar t W^3_\mu&&\gw&&\-\lover1/2 &&\-\lover1/2 &\cr
\tablerule
&b\bar bW^3_\mu && \gw&& -\lover1/2 && -\lover1/2 &\cr
\tablerule
}}\eqno{({\bf A} 4)}$$
After the electro-weak rotation one finds the standard vertex factors
$$\def\tablerule{\noalign{\hrule}}
\def\-{\hphantom-}
\vcenter{\halign{
\vrule#&\strut\quad\hfill$#$\quad\hfill&\vrule#&
\quad\hfill\strut$#$\quad\hfill&
\vrule#&\strut\quad\hfill$#$\quad\hfill&\vrule#&
\strut\quad\hfill$#$\quad\hfill&\vrule#\cr
\tablerule
&\rm Particles && g && \cv &&\-\ca&\cr
\tablerule
&t\bar t A_\mu &&\ge&& 2q_t = \lover4/3 && \-0 &\cr
\tablerule
&b\bar bA_\mu && \ge &&2q_b = -\lover2/3 && \-0 &\cr
\tablerule
&t\bar t Z_\mu&&\gz&&\-\lover1/2 -\lover4/3 \swsq&&\-\lover1/2 &\cr
\tablerule
&b\bar bZ_\mu && \gz&& -\lover1/2 +\lover2/3 \swsq && -\lover1/2 &\cr
\tablerule
}}\eqno{({\bf A} 5)}$$
\noindent
Finally the charged vertices are not affected by electro-weak mixing
$$\def\tablerule{\noalign{\hrule}}
\def\-{\hphantom-}
\vcenter{\halign{
\vrule#&\strut\quad\hfill$#$\quad\hfill&
\vrule#&\quad\hfill\strut$#$\quad\hfill&
\vrule#&\strut\quad\hfill$#$\quad\hfill&
\vrule#&\strut\quad\hfill$#$\quad\hfill&\vrule#\cr
\tablerule
&\rm Particles && g && \cv &&\-\ca&\cr
\tablerule
&e\nu \wpm_\mu &&\gw/\sqrt2&&1 && 1 &\cr
\tablerule
&t b \wpm_\mu &&\gw/\sqrt2&&1 && 1 &\cr
\tablerule
}}\eqno{({\bf A} 6)}$$
\noindent
All calculations are performed in the unitary gauge,
with propagators
$$\Delta^{\mu\nu}_f(k^2)={1\over m^2}{m^2 g^{\mu\nu}-k^\mu k^\nu\over
k^2-m^2}\eqno{({\bf A} 7)}$$
for the massive vector bosons.
All three-boson vertices have the general form
$$\hbox{3-vertex}=ig\big(g_{\mu\nu}(p_3-p_1)_\sigma
+g_{\rho\sigma}(p_1-p_2)_\mu
+g_{\mu\sigma}(p_2-p_3)_\rho \big).\eqno{({\bf A} 8)}$$
The 4-boson vertex factor is
$$ \hbox{4-vertex}=-ig^2(2g_{\mu\nu}g_{\rho\sigma}
-g_{\mu\rho}g_{\nu\sigma }
-g_{\mu\sigma }g_{\nu\rho}).\eqno{({\bf A} 9)}
$$
Only the coupling constant is different in different cases.

\endgroup

\vfill\eject

\noindent{\bf References}
\medskip
\begingroup

\parindent=-1.5em
\leftskip=1.5em
\def\ref#1{\par\leavevmode\rlap{#1)}\kern1.5em}
\def\aut#1{{\rm#1}}   \let\author =\aut
\def\tit#1{{\sl#1}}   \let\title  =\tit
\def\yea#1{{\rm(#1)}} 
\def\pub#1{{\rm#1}}   
\def\jou#1{{\rm#1}}   
\def\vol#1{{\bf#1}}   
\def\pag#1{{\rm#1}}   \let\page  =\pag

\def\nl{\hfill\break}
\let\lin\relax

\ref{1}
 \aut{Lodder J.J.},
 \tit{Towards a Symmetrical Theory of Generalised Functions},
\nl
 \jou{CWI tract }\vol{79}
 \yea{1991},
 \pub{CWI, Amsterdam}.

\ref{2}
 \aut{Lodder J.J.},
 \lin\jou{Physica}
 \vol{116A}
 \yea{1982} \pag{45, 59, 380, 392},
 \jou{Physica}
 \vol{132A}
 \yea{1985} \pag{318}.

\ref{3}
 \aut{Lodder J.J.},
 \lin\jou{Physica}
 \vol{120A}
 \yea{1983} \pag{1, 30, 566, 579}.

\ref{4}
\aut{Donoghue, J. F., Golowich E., Holstein B. R.},
\nl
\tit{Dynamics of the standard model},
\pub{Cambridge University Press},
\yea{1992}.

\ref{5} 
 \aut{Lodder J.J.},
hep-ph/9405264

\ref{6}
 \aut{Coleman, S. and Weinberg, E.},
\jou{Phys. Rev.}
\vol{D7}
\yea{1973}
\page{1888}

\ref{7}
 \aut{Weinberg, S.},
\jou{Phys. Lett.}
\vol{82B}
\yea{1979}
\page{387}

\ref{8} 
 \aut{Lodder J.J.},
hep-ph/9405265

\ref{9}
 \aut{Lodder J.J.},
 hep-ph9606nnn

\endgroup

\bye